\documentclass[
aps,%
12pt,%
final,%
notitlepage,%
oneside,%
onecolumn,%
nobibnotes,%
nofootinbib,%
superscriptaddress,%
noshowpacs,%
centertags]%
{revtex4}

\begin{document}
\selectlanguage{english}





\title{GIANT PULSES --- THE MAIN COMPONENT OF THE RADIO EMISSION OF
THE CRAB PULSAR}

\author{\firstname{M.~V.}~\surname{Popov}}
\affiliation{Astro Space Center, Lebedev Physical Institute, Moscow, Russia}

\author{\firstname{V.~A.}~\surname{Soglasnov}}
\affiliation{Astro Space Center, Lebedev Physical Institute, Moscow, Russia}

\author{\firstname{V.~I.}~\surname{Kondrat'ev}}
\affiliation{Astro Space Center, Lebedev Physical Institute, Moscow, Russia}

\author{\firstname{S.~V.}~\surname{Kostyuk}}
\affiliation{Astro Space Center, Lebedev Physical Institute, Moscow, Russia}

\author{\firstname{Yu.~P.}~\surname{Ilyasov}}
\affiliation{Pushchino Radio Astronomy Observatory, Astro Space Center,
Lebedev Physical Institute, Pushchino, Russia}

\author{\firstname{V.~V.}~\surname{Oreshko}}
\affiliation{Pushchino Radio Astronomy Observatory, Astro Space Center,
Lebedev Physical Institute, Pushchino, Russia}

\received{May 20, 2005}
\revised{July 6, 2005}

\begin{abstract}
The paper presents an analysis of dual-polarization observations of the Crab
pulsar obtained on the 64-m Kalyazin radio telescope at 600~MHz with a time
resolution of 250~ns. A lower limit for the intensities of giant pulses is
estimated by assuming that the pulsar radio emission in the main pulse and
interpulse consists entirely of giant radio pulses; this yields estimates of
100~Jy and 35~Jy for the peak flux densities of giant pulses arising in the
main pulse and interpulse, respectively. This assumes that the normal radio
emission of the pulse occurs in the precursor pulse. In this case, the longitudes
of the giant radio pulses relative to the profile of the normal radio emission
turn out to be the same for the Crab pulsar and the millisecond pulsar
B1937$+$21, namely, the giant pulses arise at the trailing edge of the profile
of the normal radio emission. Analysis of the distribution of the degree of
circular polarization for the giant pulses suggests that they can consist of a
random mixture of nanopulses with 100\% circular polarization of either sign,
with, on average, hundreds of such nanopulses within a single giant pulse.
\end{abstract}

\maketitle

\section{INTRODUCTION}

Giant pulses are short-time-scale flares of radio emission whose peak flux
densities exceed the peak flux density in the mean pulse profile of the
pulsar by a factor of hundreds or thousands. Although there have recently
been a number of publications reporting the detection of strong, brief pulses
in several pulsars~[1\mbox{--}5], detailed studies of the properties of giant
radio pulses have been presented for only two pulsars: the Crab pulsar
(B0531$+$21; see references in~[6,~7]) and the millisecond pulsar
B1937${+}$21~[8--12].

\section{OBSERVATIONS AND REDUCTION}
\label{obs:Popov_n}

The observations considered here were carried out on November 25--26, 2003
at 600~MHz on the 64-m Kalyazin radio telescope. The duration of the observing
session processed was three hours. The observations were obtained in the
framework of an international program of multi-frequency studies of the
properties of the Crab pulsar's giant pulses. Other telescopes participating
in this program include the 100-m Effelsberg radio telescope (8350~MHz), 76-m
Lovell Telescope at Jodrell Bank (1400~MHz), the Westerbork Radio Synthesis
Telescope (1200~MHz), the Large Phased Array and DKR-1000 radio telescope of
the Pushchino Radio Astronomy Observatory (111~MHz), and the T-shaped UTR-2
radio telescope in Khar'kov (23~MHz). In addition, simultaneous optical
observations were obtained on the 6-m telescope of the Special Astrophysical
Observatory and the 2.9-m telescope at La~Palma. The MAGIC and HESS gamma-ray
telescopes in La~Palma and Namibia, which detect Cerenkov radiation in the
upper layers of the atmosphere due to the passage of high-energy gamma-rays,
also participated. We present here only an analysis of the Kalyazin
observations at 600~MHz. A joint analysis of the multi-frequency observations
will be presented in other publications.

\begin{figure*}[t!]
\vskip -30mm
\hskip -10mm\includegraphics[scale=1.]{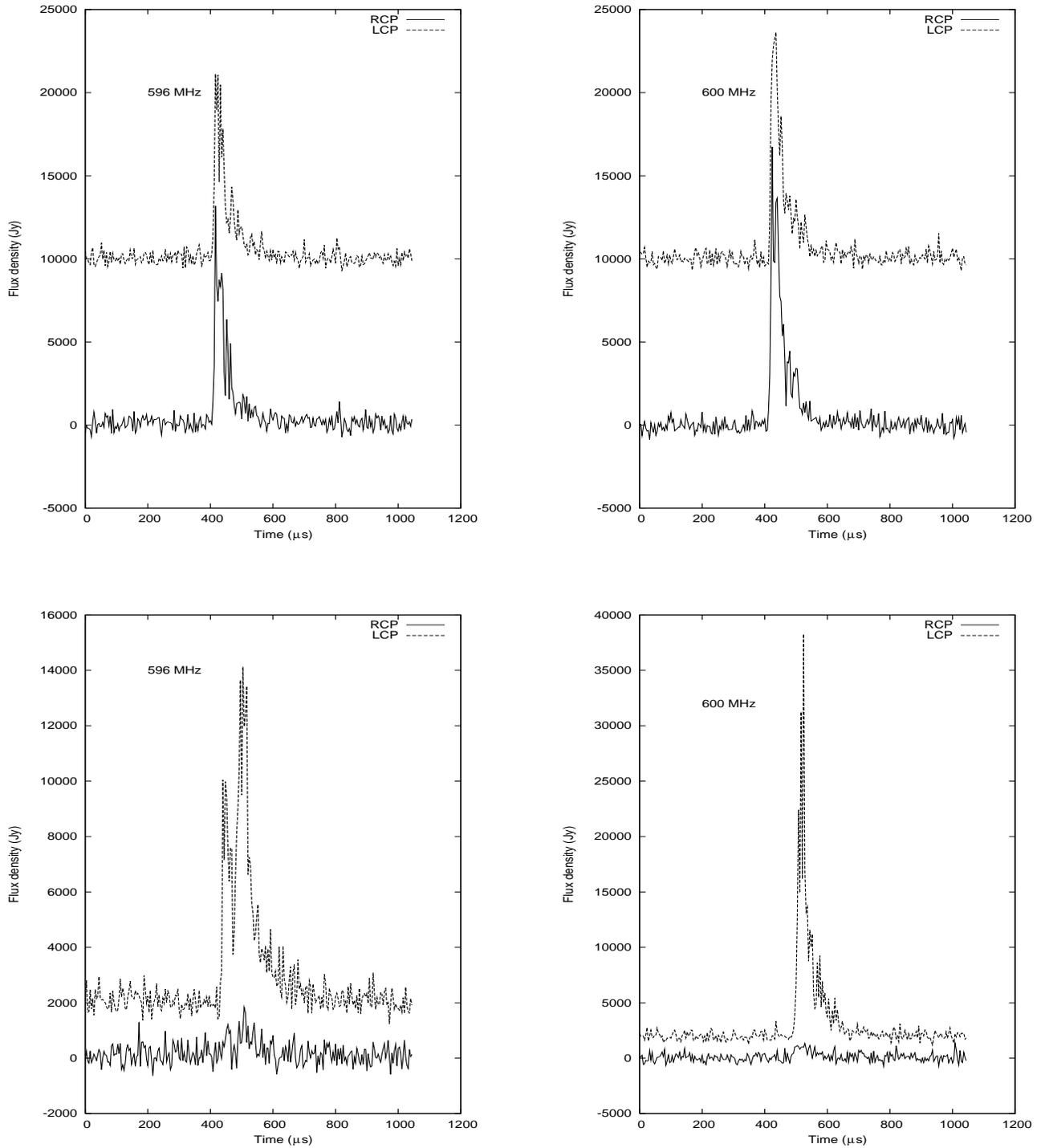}
\setcaptionmargin{5mm}
\onelinecaptionsfalse  
\vskip -30mm
\caption{Two strong giant pulses after averaging to 8~ms. The dashed and
solid curves show the signals in left- and right-circular polarization,
respectively. The observed shape of the pulses is due to scattering of the
radio waves on inhomogeneities in the interstellar plasma.\hfill}\vspace*{-5pt}
\end{figure*}

The Kalyazin radio telescope received radio emission in two channels,
sensitive to left- and right-circular polarization (LCP and RCP). Each
polarization channel recorded two frequency bands (upper and lower
sideband), each with a width of 4~MHz, with the central frequency being
600.0~MHz. Thus, the time resolution of the observations was 250~ns. The
data were recorded on a video cassette with two-bit sampling using an S2
recording system. The data were played back after the observations at the
Astro Space Center of the Lebedev Physical Institute using the specialized
S2-RDR system developed in the Astrophysical Data-Reduction Department of
the Astro Space Center. We removed the influence of the dispersion of the
radio waves using the method of pre-detection dispersion compensation~[13].
The reduction methods used are described in~[14]. The procedure used to
reconstruct the signal is very computer-intensive: 10 minutes of observational
data were processed in 20~hours. When reconstructing the signal, we used the
emission measure of 56.757, taken from the monthly ephemerides of the Jodrell
Bank Observatory~[15].

\section{DETECTION CRITERIA}
\label{giant:Popov_n}

Broadening of the pulse due to dispersion in the interstellar plasma
(by about 8.5~ms in our case) was fully corrected for using the method of
pre-detection dispersion compensation. However, scattering on inhomogeneities
in the interstellar plasma also significantly broadens the pulse. As we will
show in Section~4, the effective width of the pulse due to scattering was
60~$\mu$s for our session. Therefore, to increase the sensitivity of the search
for giant pulses, we averaged the reconstructed signal over 32~$\mu$s (256
points). Since it is well known that giant radio pulses from the Crab pulsar
are observed only in narrow longitude intervals corresponding to the positions
of the main pulse and interpulse, we searched for radio flares only in these two
longitude ``windows,'' synchronized with the pulsar period. The width of each
window was 960~$\mu$s, or 30 points of the averaged recording. Events were taken
to represent giant pulses if the amplitude of the signal in one of the windows
exceeded $5\,\sigma$ relative to the mean level in at least one of the four
channels. We also required that the amplitude at that same point in one of the
three remaining channels (allowing for the dispersion delay between channels)
exceeded $3.5\,\sigma$. After averaging over 256 points, the distribution of
signal amplitudes was close to a normal distribution, for which the
probabilities of exceeding the $5\,\sigma$ and $3.5\,\sigma$ levels are
$2.87\times 10^{-7}$ and $2.32\times 10^{-4}$, respectively. The total
probability of a random realization of our detection criteria is then
$8\times 10^{-10}$. In each pulsar period, 60 measurements are made, and
$2\times 10^{7}$ are made over three hours of observations, so that the
expected number of false giant pulses is only 0.015. Thus, all the detected
giant pulses are real. Over the three hours of observations, we detected
4287 giant pulses, of which 3802 were at the longitudes of the main pulse
and 485 at the longitudes of the interpulse.

\begin{figure}[t!]
\includegraphics[scale=0.5]{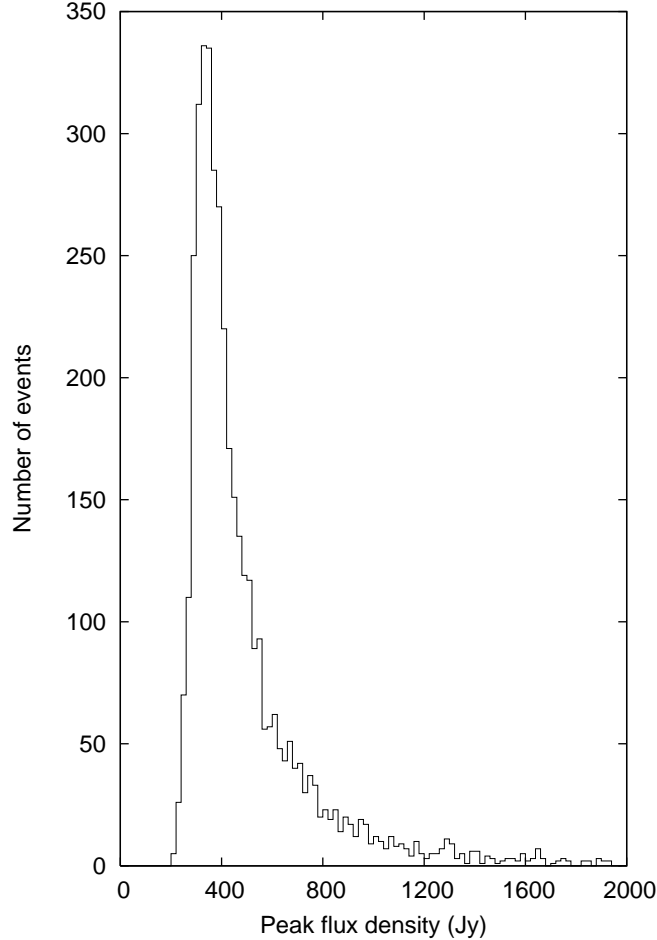}
\setcaptionmargin{5mm}
\onelinecaptionsfalse  
\caption{Distribution of the peak flux densities of the detected giant
pulses.\hfill}\vspace*{-5pt}
\end{figure}

Figure 1 shows examples of strong giant pulses. The flux-density scale is
established in accordance with the rms deviation, which is 1270~Jy, and is
determined primarily by the contribution of the radio emission of the Crab
nebula. The basis for this estimation is given in~[16]. For our subsequent
analysis, we must know the effective threshold peak flux density of the giant
pulses detected according to the described criteria. This value was determined
from the distribution of the peak flux densities of the detected giant pulses
shown in Fig.~2. The peak flux density was determined by averaging over
all frequencies and polarizations. The distribution in Fig.~2 shows a sharp
cutoff at a flux density of about 300~Jy, and this value was adopted for
the minimum peak flux density for the giant pulses detected in our analysis.

\section{SCATTERING PROFILE}
\label{prof_scat:Popov_n}

\begin{figure*}[t!]
\vskip -70mm
\includegraphics[scale=0.95]{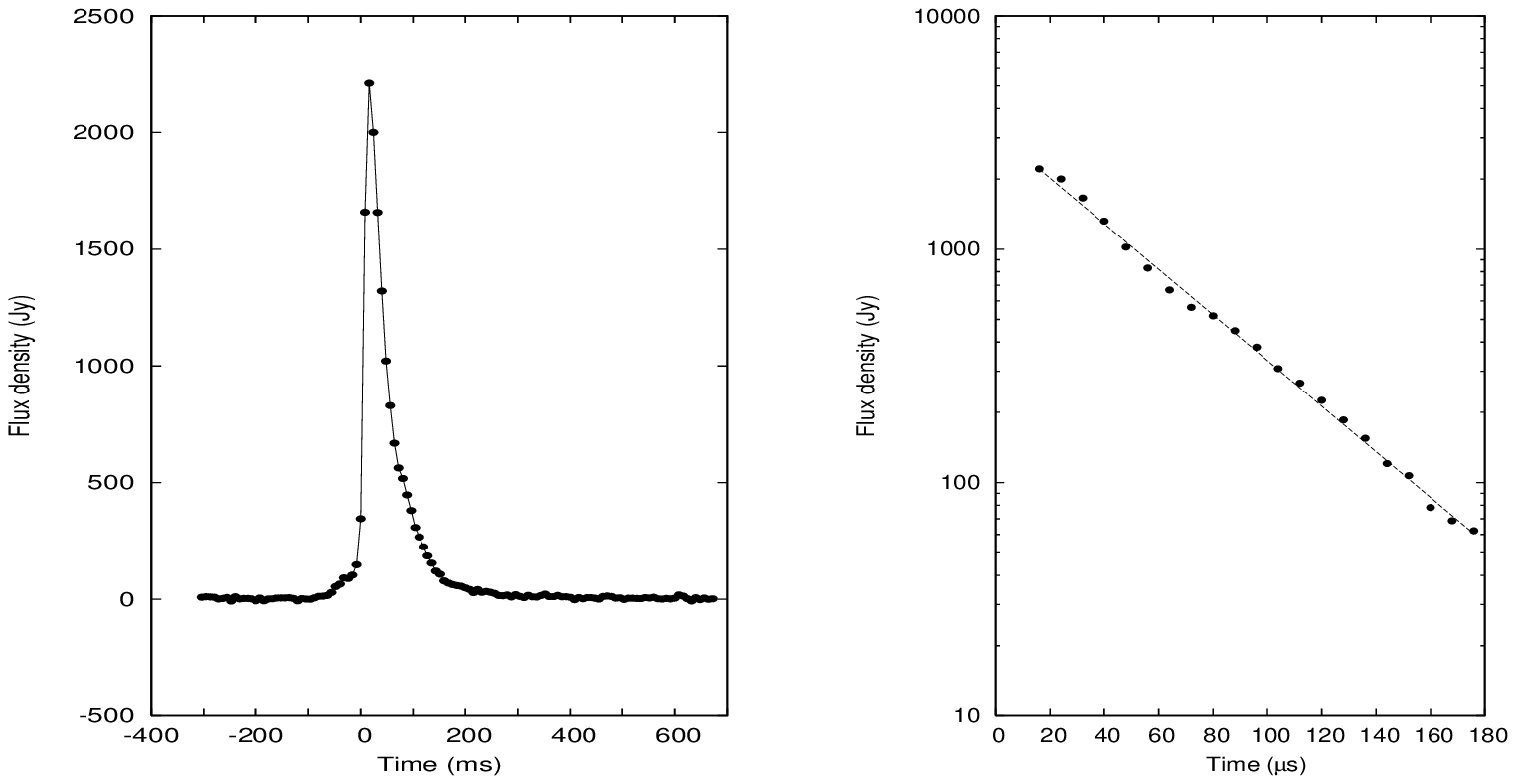}
\setcaptionmargin{5mm}
\onelinecaptionsfalse  
\vskip -80mm
\caption{The scattering profile of the giant pulses (left) and an exponential
approximation to the tail of the profile (right). The exponential time constant
is 45~$\mu$s.\hfill}
\end{figure*}

\begin{figure}[t!]
\includegraphics[scale=0.5]{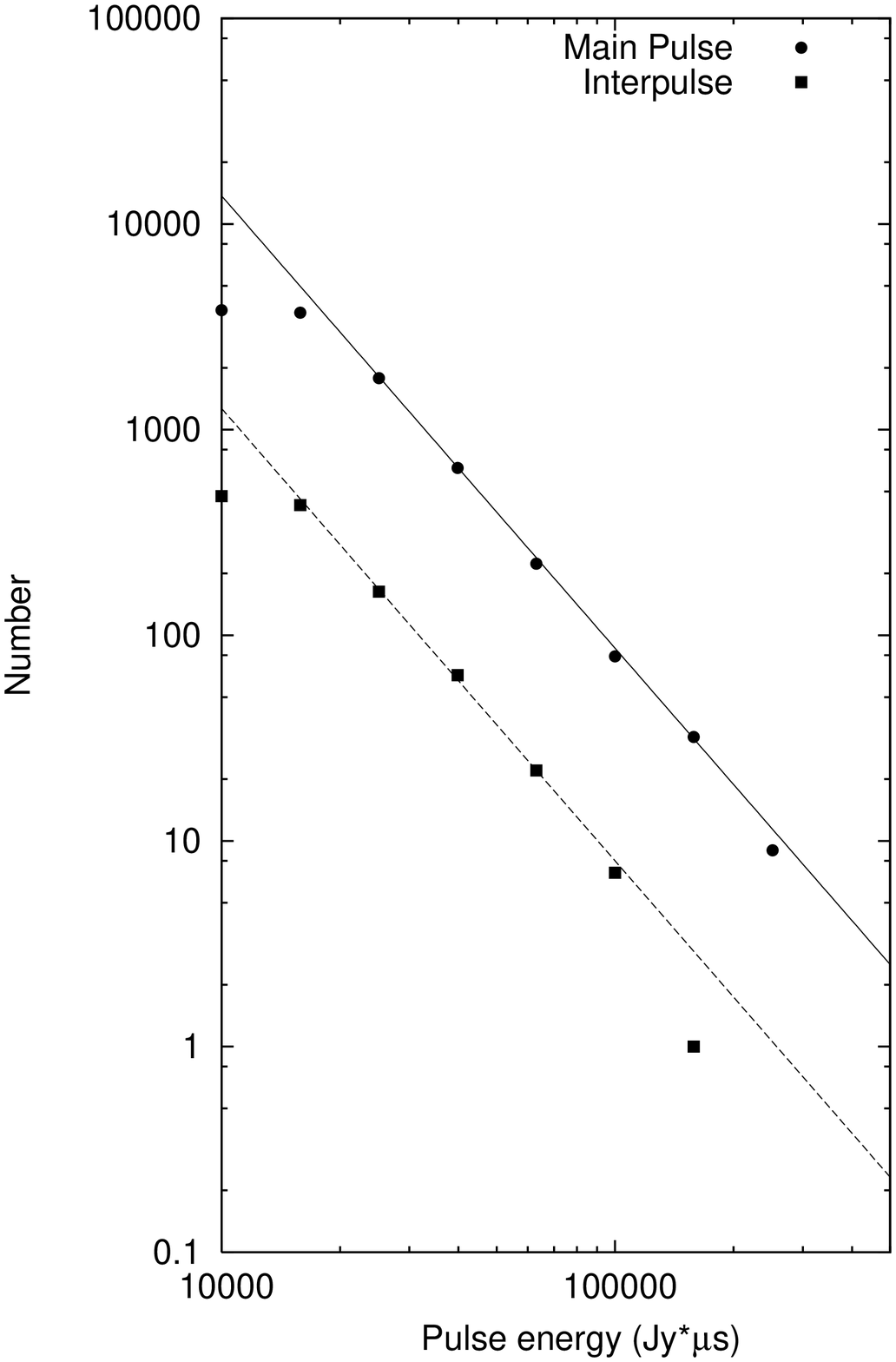}
\setcaptionmargin{5mm}
\onelinecaptionsfalse  
\caption{Integrated energy distribution of the giant pulses. The circles and
squares show data for giant pulses detected at the longitudes of the main
pulse and interpulse, respectively. The straight lines correspond to a power
law with index ${-}2.2$.
\hfill}
\end{figure}

\looseness=-2 The shape of the detected giant radio pulses is fully
determined by scattering on inhomogeneities of the interstellar plasma, as
we can clearly see in Fig.~1. The giant pulses have a steep front and an
exponential tail. To determine the effective widths of the giant pulses, we
constructed a mean giant-pulse profile by averaging all the detected pulses
with peak flux densities greater than 800~Jy. During the averaging, the pulses
were equalized according to the point at the leading front, for which the
amplitude of the deviation from the mean value outside the pulse was less than
30$\%$ of the maximum amplitude of the given pulse (this equalization was
carried out for the signal averaged over a time of 8$\mu$s). The resulting
mean giant-pulse profile is shown in the right-hand part of Fig.~3. The left-hand
part of this figure shows the time dependence of the intensity for the trailing
edge of the pulse on a half-log scale. We can see that the observed points
are approximated well by a linear dependence, and that the shape of this
part of the profile corresponds to an exponential decay,
$S=S_pe^{-t/\tau}$, with the characteristic time $\tau = 45 \pm{5}$~$\mu$s.
Taking into account the leading front of the pulse, we adopted for the
effective width of the giant-pulse profile $W_e = 60~\mu$s. This width was
used to calculate the energy of the detected giant pulses $E=S_pW_e$, where
$S_p$ is the peak flux density.

\section{ENERGY DISTRIBUTION}
\label{energy-dist:Popov_n}

\begin{figure*}[t!]
\vskip -30mm
\includegraphics[scale=0.95]{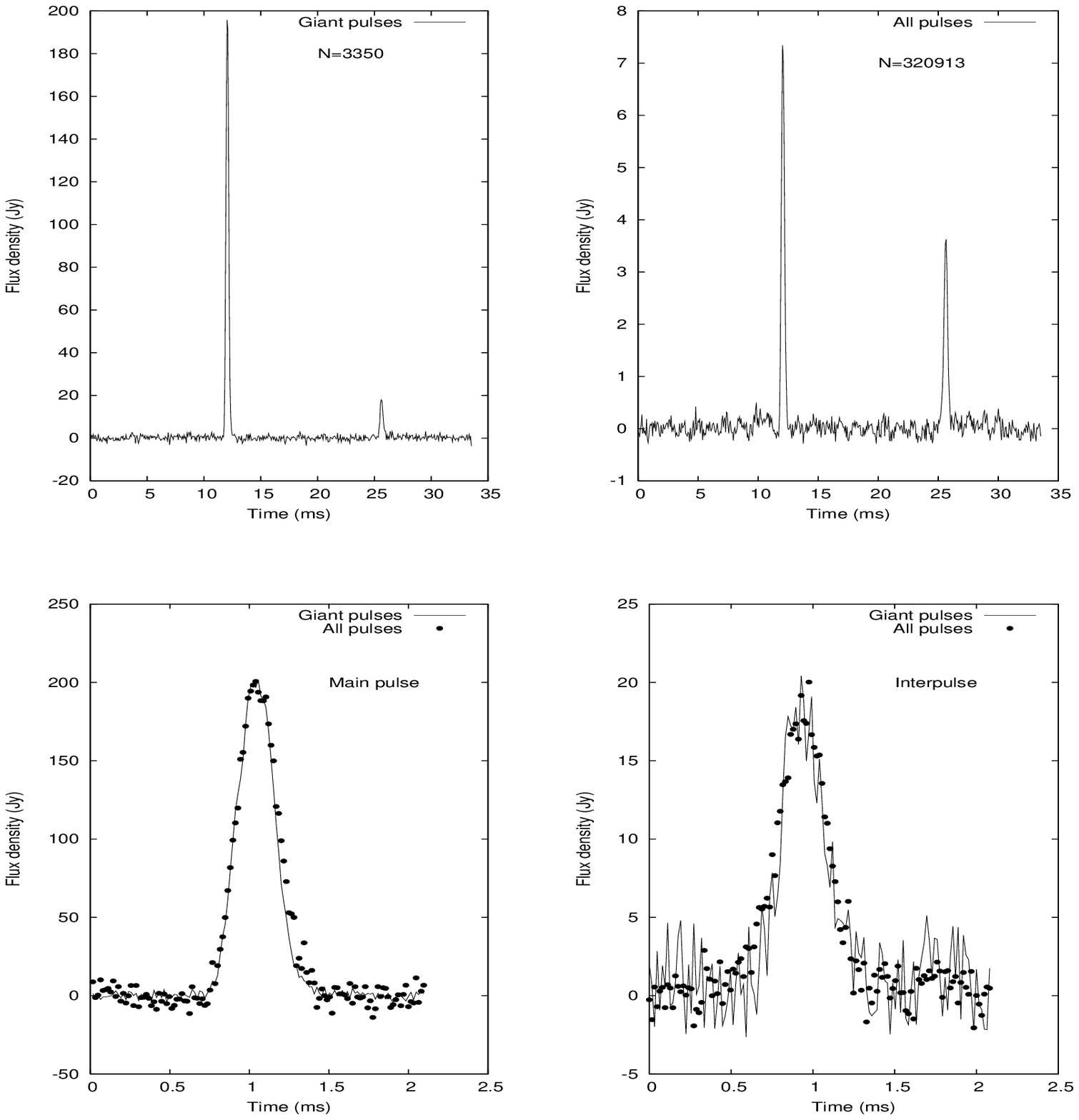}
\setcaptionmargin{5mm}
\onelinecaptionsfalse  
\vskip -40mm
\caption{Upper plots: mean profiles obtained by averaging 3350 giant
pulses (left) and by averaging all 320\,913 pulses (right). Lower plots:
comparison of the profile shapes after the mean profiles have been scaled
to have the same amplitude.\hfill}
\end{figure*}

\begin{figure}[t!]
\includegraphics[scale=0.5]{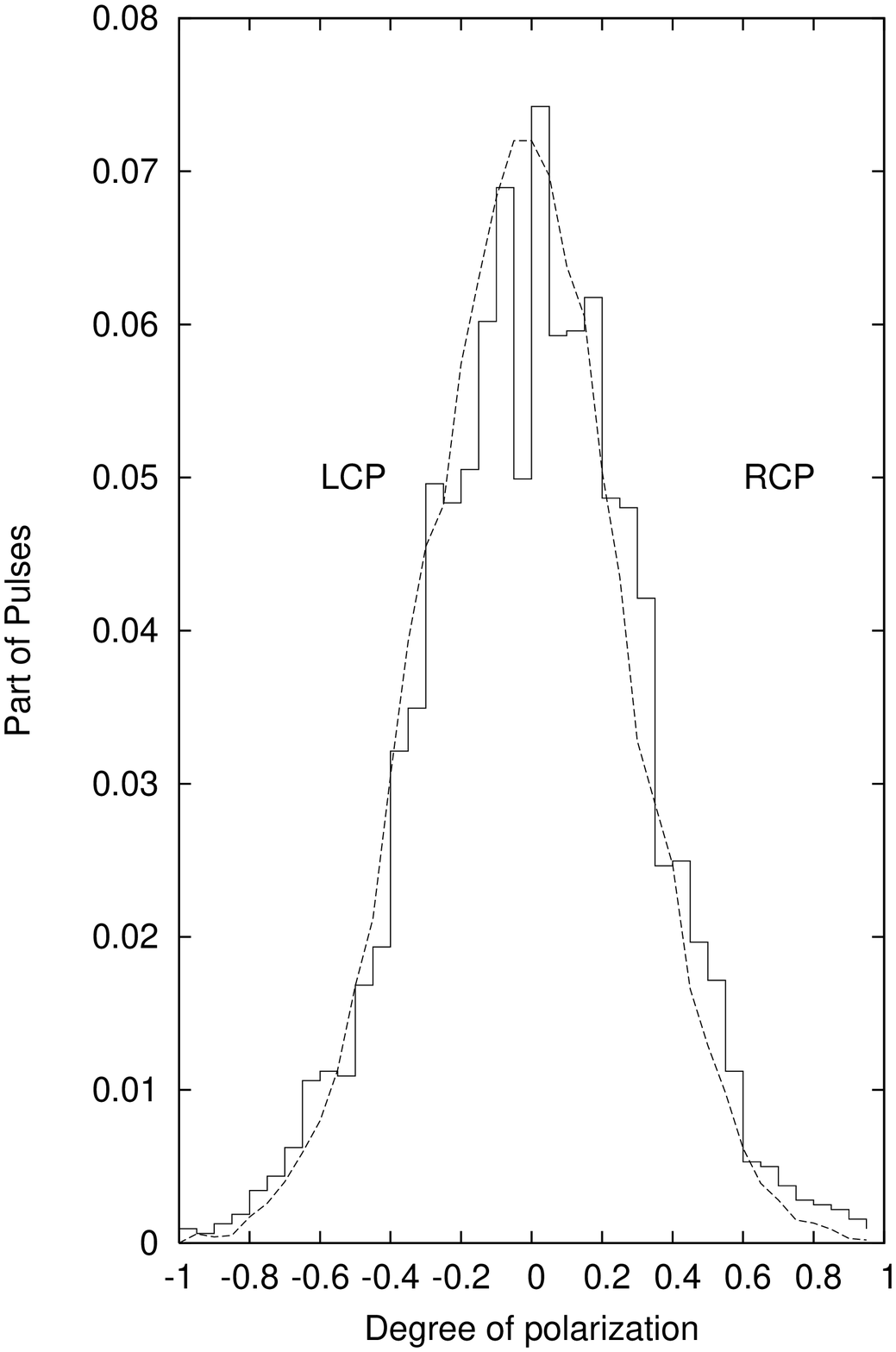}
\setcaptionmargin{5mm}
\onelinecaptionsfalse  
\caption{Distribution of the degree of circular polarization for the
giant pulses. The dashed curve shows the results of our simulations (see text
for more information).\hfill}
\end{figure}

Figure~4 shows the distributions of the integrated energy of the detected giant
pulses separately for pulses at the longitudes of the main pulse and interpulse.
Near the mean energies, both distributions are approximated well by power
laws, ${N(E) \propto E^{-\alpha}}$, with the same index, $\alpha=2.20\pm 0.06$.
Deviations from this behavior at small energies are due to the incompleteness
of the pulse detection, while deviations at large energies are due to
insufficient statistics. The most representative studies of the statistics of
giant pulses from the Crab pulsar are those of Lundgren et al.~[17], who
detected about 30\,000 giant pulses over several days of observations with the
43-m Green Bank radio telescope at 800~MHz. The values of the index $\alpha$
they derived were different on different days, and were contained in the
interval 3.06--3.36. Cordes et al.~[18] obtained the index $\alpha=  2.3$
for the distribution of the peak flux densities of giant pulses observed at
430~MHz, which is much closer to our value derived at 600~MHz.

\section{GIANT PULSES AND THE MEAN PROFILE OF THE PULSAR}
\label{av-prof:Popov_n}

Figure 5 compares the mean profiles of B0531$+$21 obtained by
averaging all the pulses detected over the observation period (${N=320\,913}$)
and averaging only the detected giant pulses ($N=3350$). We can clearly see
that the mean profile for the giant pulses has a higher signal-to-noise ratio,
although the total profile includes a hundred times as many
individual pulses. This same conclusion was reached by Cordes et al.~[18].
Moreover, the shapes of the two profiles (lower part of Fig.~5) are surprisingly
similar, as was earlier pointed out by Friedman and Boriakoff~[19]. All this
suggests that the pulsar emission in the main pulse and interpulse consist
entirely of giant radio pulses. We suggest that the ``normal'' radio emission
of the pulsar is concentrated in the precursor pulse, which is barely
distinguishable in our observations (upper right plot in Fig.~5). In this
case, the longitudes of the giant pulses in the mean profile of B0531${+}$21
become identical to the longitude distribution for the giant pulses of the
millisecond pulsar B1937$+$21, for which all the giant pulses arise at the
trailing edges of the main pulse and interpulse~[8\mbox{--}10]. Friedman and
Boriakoff~[19] also suggest the possible presence of a precursor pulse near
the interpulse of the Crab pulsar, providing additional evidence in support
of our hypothesis.

Assuming that the radio emission in the main pulse and interpulse consists
only of giant pulses and that the energy distribution of these pulses follows
a power law with index ${-}2.2$ down to some lower limit, we can determine this
limit by comparing the total energy in the detected giant pulses, which have a
known detection limit, and the total energy in all the pulses observed in the
session. This yields for the giant pulses at the longitudes of the main pulse
and interpulse lower limits for the peak flux density of 105~Jy and 25~Jy,
respectively. Let us now estimate the total number of potentially observable
giant pulses at 600~MHz. If we change these estimates slightly to the values
100 and 35~Jy, which lie within their uncertainties, the total number of
potentially observable giant pulses at the longitude of the main pulse and
interpulse become the same, equal to 42\,000 pulses over the three hours of
observations, which corresponds to a mean rate of appearance of the giant
pulses of four per second. Using the fact that the giant pulses can be
observed only within narrow longitude intervals corresponding to the observed
width of the profiles of the main pulse and interpulse (about 0.5~ms), we
can estimate the total mean rate of appearance of giant pulses, which is
approximately 270 pulses per second, or about 10 giant pulses per rotation
of the neutron star, for the longitudes of both the main pulse and interpulse.

Lundgren et al. [17] estimated the minimum threshold peak flux density for
the giant pulses from B0531$+$21, obtaining about 50~Jy at 800~MHz without
distinguishing between giant pulses corresponding to the main pulse and
interpulse. They assumed the presence of weak, continuously arising
micropulses with flux densities of 1--2~Jy in addition to the giant pulses.
We cannot refute this suggestion, but consider it to be unnecessary.

\section{POLARIZATION OF THE GIANT PULSES}

The two giant pulses presented in Fig.~1 as examples display completely
different polarization characteristics: the pulse in the upper plots has equal
intensities in RCP and LCP, while the pulse in the lower plots is almost 100\%
left-circularly polarized. Little has been published about the polarization
properties of individual giant pulses. In their study of giant pulses from
the millisecond pulsar B1937$+$21, Cognard et al.~[8] show that giant pulses
with high degrees of either LCP or RCP frequently appear. Popov et al.~[11]
also reported the detection of giant pulses from B1937$+$21 with high degrees
of circular polarization of both signs at 600~MHz. Hankins et al.~[7] showed
that nanopulses detected in the structure of giant pulses from B0531$+$21 have
high degrees of circular polarization, with individual nanopulses separated
in time by several microseconds sometimes displaying opposite signs of
circular polarization.

Returning to our 600-MHz observations, we note that the pulse with strong
circular polarization shown in Fig.~1 is the only such example among the
strong giant pulses. The solid step-wise line in Fig.~6 shows the observed
distribution of the circular polarization for all detected giant pulses with
peak flux densities exceeding 500~Jy. We calculated the degree of polarization
as $\dfrac{R-L}{R+L}$, where $R$~and~$L$ are the peak flux densities in
RCP and LCP averaged over the two frequency channels. We can see from Fig.~6
that pulses with high degrees of polarization are not rare, although the
vast majority of the pulses have degrees of polarization in the range
${-}0.4\ldots{+}0.4$.

\looseness=-2 Based on the results of Hankins et al.~[7], we supposed that all
giant pulses consist of nanopulses with 100\% circular polarization of one or
the other sign, and carried out numerical simulations of the expected
circular-polarization distribution for events under this assumption. In the
simulations, individual nanopulses were generated as the sum of the squares
of two components, which we arbitrarily call the real and imaginary components
of the nanopulse. The components were taken from a normal distribution with
a zero mean and unit dispersion. The sign of the polarization of the nanopulse
was determined by the sign of the imaginary component.  The amplitude
distribution for the generated nanopulses corresponds to a $\chi^2$ distribution
with two degrees of freedom, as should be the case for a detected signal without
averaging. For each simulated giant pulse, we specified the number of constituent
nanopulses, again calculating the degree of polarization as $\dfrac{R-L}{R+L}$,
where $R$ and $L$ are the sums of the amplitudes of the nanopulses with RCP
(a positive imaginary component) and LCP (a negative imaginary component),
respectively. The simulations were carried out for various numbers of
constituent nanopulses from 10 to 1000. We ran 10\,000 trials in order to
construct the distribution. The dashed line in Fig.~6 shows the simulated
distribution for the case of 100 nanopulses making up each giant pulse,
which is the model distribution that best agrees with the observed distribution.

\section{DISCUSSION}
\label{disc:Popov_n}

\looseness=-2 Our suggestion that the radio emission of the Crab pulsar
at the longitudes of the main pulse and interpulse consists entirely of
giant radio pulses removes one substantial difference between the properties
of the giant pulses of the Crab pulsar and the millisecond pulsar B1937$+$21.
It was immediately noted for the latter pulsar that the giant pulses arise
outside the mean profile for the normal radio emission, at the trailing edges
of the profiles of the main pulse and interpulse~[9, 10]. We propose that the
normal radio emission of the Crab pulsar is generated at the longitudes of
the precursor pulse preceeding the main pulse and interpulse, so that the
giant pulses making up these components are localized at the trailing edges
of the profiles corresponding to the normal radio emission, as in the case of
the millisecond pulsar B1937$+$21. This unification of the properties of the
giant radio pulses of these two pulsars could be very important for our
understanding of the origin of the giant pulses.

Another appreciable difference in the parameters of the giant pulses observed
from the millisecond pulsar B1937$+$21 and the Crab pulsar is their
characteristic duration. The analysis of Hankins~[20], carried out at
4.9~GHz, where scattering on inhomogeneities in the interstellar plasma does
not appreciably affect the pulse profile, shows that the intrinsic durations of
the giant pulses from the Crab pulsar are typically several microseconds. The
giant pulses from B1937${+}$21 have been unresolved in all published studies.
Soglasnov et al.~[10] showed that the intrinsic durations of the giant pulses
from this pulsar are less than 15~ns. However, Hankins et al.~[7] showed that
the giant pulses generated by the Crab pulsar include some that consist entirely
of unresolved nanopulses with durations of less than 2~ns, with these nanopulses
often displaying 100\% circular polarization. Our analysis in Section~7 indicates
that the circular-polarization distribution for the giant pulses is consistent
with the hypothesis that these pulses consist of, on average, hundreds of
nanopulses, each with 100\% LCP or RCP. Thus, the difference in the durations
of the giant radio pulses from the millisecond pulsar and the Crab pulsar
is not important. Namely, this difference corresponds only to differences in
the number of nanopulses comprising each giant pulse: no more than a few
dozen in the case of the millisecond pulsar and, on average, hundreds in the
case of the Crab pulsar. We suggest that the nanopulses making up the giant
pulses from the Crab pulsar and millisecond pulsar have the same nature.

\section{ACKNOWLEDGMENTS}

The authors thank K.G.~Belousov and A.V.~Chibisov for providing the
operational S2-RDR play-back system. This work was supported by the Russian
Foundation for Basic Research (project code 04-02-16384) and the basic
research program of the Presidium of the Russian Academy of Sciences on
``Non-stationary Phenomena in Astronomy.''

\end{document}